\documentstyle[aps,preprint]{revtex}
\draft

\begin{document}

\newcommand{\uprule}{\end{multicols}
\noindent \vrule width3.375in height.2pt depth.2pt 
\vrule height.5em depth.2pt \hfill \widetext }
\newcommand{\downrule}{\indent \hfill \vrule depth.5em height0pt 
\vrule width3.375in height.2pt depth.2pt 
\begin{multicols}{2} \narrowtext}	
	
\include{psfig}

\draft \title{\bf Quantum computing in a Spin ordered phase}

\author{F. de Pasquale $^{a,b}$ and S.M. Giampaolo $^{a}$}

\address{\mbox{$^a$ 
Departement of Physics University of Rome 
"La Sapienza", P-le A.Moro 5, 00185 Rome Italy }} 
\address{\mbox{$^b$ INFM section of Rome}}

\date{Feb. 01 2001} \maketitle

\begin{abstract}

We show that quantum computation can be performed in a system at thermal 
equilibrium if a spontaneous symmetry breaking occurs.
The computing process is associated to the time evolution of the statistical 
average of the qubit coherence operator. This average defines logical states 
which evolve under the action of impulsive gate perturbations. 
Non trivial quantum coherence properties appear in the ordered phase 
characterized by a spontaneous symmetry breaking. Qubits are associated to 
spins and interaction with the lattice vibrations is approximated by an 
effective spin-spin interaction. A finite decoherence time is obtained only 
if spin correlations are taken into account in the framework of the Ising 
model as corrections to the Mean Field approximation. Decoherence is slowed 
down as the order parameter saturates.
\end{abstract}
%\begin{multicols}{2} 
%%%%%%%%===================introduzione===================================

In the quantum computing process\cite{Rieffel} information is stored in 
qubits i.e. in a  quantum state of a two levels system. Coupling with the 
environment 
produces an ``entanglment'' of the qubits with the degrees of freedom of the 
environment. This ``entanglement'' is the origin of a progressive loss 
of information, a process known as decoherence. Hence, systems at thermal 
equilibrium have not been considered suitable for a computing process. We  
show that quantum computation is however possible even in thermal equilibrium 
in presence of a spontaneous symmetry breaking. 
Logical states of computer will be associated to suitable thermal averages. 
The 0-logical state is associated to thermal equilibrium and 1-logical state 
to local excitation determined by an impulsive perturbation. Decoherence is 
then associated to the local excitation relaxation toward equilibrium. We 
shall see on a specific model that relaxation vanishes in the Mean Field 
Approximation (MFA) and is slowed down as magnetization saturates when 
corrections to MFA are taken into account.   

For sake of simplicity we work with one of the most simple system which 
exhibits symmetry breaking: the Ising model 
$H = -\sum_{i,j} J_{i,j}S_i^{z} S_j^{z}$. 
This model is related to the more familiar spin - boson model\cite{LDV} where 
spins interact with environment vibrations. If up and down local spin states
are degenerated in energy, lattice degrees of freedom can be eliminated by a 
canonical transformation. Ising model is then obtained as an approximation if 
lattice modes polarized in the spin quantization direction have smaller 
frequencies than perpendicular ones.  
Another physical realization is given by quantum dots interacting with the
oscillation modes of the substrate. In this case the two states 
( $|\uparrow>$ and $|\downarrow>$ ) can be associated to the presence or 
absence of an extra electron. In Coulomb blockade condition, i.e. when the 
two states are degenerated in energy, the coupling of quantum dots with 
phonons gives rise to an effective interaction between spins equivalent to the
Ising model. Another possible physical realization of this model is a system 
of bosonic neutral atoms localized in the sites of a lattice interacting with 
repulsive interaction. Possible realization of such a system in an optical 
lattice has been discussed in 
\cite{demartini1,demartini2,brennen}.  

We define the i-th qubit as the state 
$|\theta_i,\phi_i> = \cos(\frac{\theta_i}{2})|\uparrow_i>+ 
e^{i\phi_{i}} \sin(\frac{\theta_i}{2})|
\downarrow_i> $ associated to the corresponding site of a regular array 
( i=0,..,N-1 ). The local quantum coherence operator is the projection 
operator on a given qubit state.
\begin{equation}
\pi_i(\theta,\phi) = |\theta_i,\phi_i><\theta_i,\phi_i|
\end{equation}
The projection operator in terms of spin operators reads as
\begin{equation}
\pi_i(\theta,\phi) = \frac{1}{2} + n_i S_i
\end{equation}
where $n$ is the unitary vector associated to the $\theta$, $\phi$ 
direction and $S$ is the vector spin operator. The statistical average of the 
quantum coherence operator is the qubit coherence probability (QCP) 
\begin{equation}
\label{simpty}
\rho_i = \frac{Tr(e^{-\beta H}\pi_i)}{Tr(e^{-\beta H})}
\end{equation}
$\rho_i$ is the probability of finding the spin at the i-th site, in 
the state $|\theta_i, \phi_i>$, when the system is in equilibrium with an 
external bath. It is immediately seen from (\ref{simpty}) that as a 
consequence of the spin inversion symmetry of the Ising Hamiltonian 
$H(\{S_{i}\}) = H(\{-S_{i}\})$ we have $\rho_i=\frac{1}{2}$. All 
directions $\theta_i,\phi_i$ are then equally probable and coherence, as 
expected,is completely lost at equilibrium. 
Thus a ``symmetric'' thermodynamic state is obviusly useless as a reference 
state in a computation process. However, if a 
spontaneous symmetry breaking occurs below a critical temperature ($T_c$), 
we obtain a non trivial QCP. In our case spin inversion  
symmetry breakdown implies $m_i = <S_i^{z}> \neq 0 $ ( angular 
brackets stand for thermodynamic average ). As a consequence the QCP becomes
\begin{equation}
\label{density}
\rho_i=\frac{1}{2}+m_i \cos \theta_i
\end{equation}  
We see in (\ref{density}) that coherence is partially preserved at equilibrium
even if the dependence on the phase $\phi_i$ is lost. In the states space the 
direction $\theta_i=0$ is the most probable and the average of any even 
function of $\theta_i$ does not vanish\cite{nota}. 
Hence, the thermodynamic state associated to 
the ordered spin phase is suitable as the reference state of a computing 
process. It is convenient to consider a reduced QCP as the difference between
the ordered and disordered phase QCPs 
$\widetilde{\rho}_i = <\widetilde{\pi}_i>=< \pi_i-\frac{1}{2}>$. We find 
convenient 
to identify the 0-logical state ( 1-logical state ) of the i-th site with the 
dependence of $\widetilde{\rho}_i$ on $ \cos \theta_i $ $(\sin \theta_i)$.
In other words different logical states are associated with different 
properties of a quantity related to the statistical average of the quantum 
coherence instead of different quantum states. 

Universal quantum computer needs 1-qubit and 2-qubits 
gates\cite{Divincenzo,Barenco}. The first is 
a rotation of the 0-logic state in a superposition of 0 and 1 logical states. 
For this reason we need a perturbation that modifies the local QCP 
introducing a term with depends on $\sin \theta_i$.
The second modifies the two sites state introducing an 
entanglement between the two logical states. In the corresponding 2-qubits QCP
a $\sin\theta_i\sin\theta_j$ will appear as a result of the perturbation 
acting on a two sites 0 logical initial state. 

For sake of simplicity we consider only impulsive perturbations.
\begin{equation}
H_{t} =  H + \epsilon\delta(t-0^{+})H_I 
\end{equation}
After the action of a gate the QCP is given by
\begin{equation}
\widetilde{\rho}_i(t) = \frac{Tr(e^{-\beta H}\widetilde{\pi}_i(t))}
{Tr(e^{-\beta H})}
\end{equation}
Where $\widetilde{\pi}_i(t)$ is the reduced quantum coherence operator in the
Heisenberg representation.

As 1-bit gate we consider a perturbation perpendicular to the z direction 
acting on a given site i=0. We choose to work with 
\begin{equation}
H_I=S_0^x
\end{equation}
If the Ising model is associated with localized particles ( quantum dots, 
optical lattices, etc. ), the perturbation could be obtained by exchanging 
particles with a reservoir while in the spin case a magnetic field\cite{Loss} 
could be used. The new logical state is characterized as 
\begin{equation}
\label{1bitgate}
\widetilde{\rho}_0(t)  = \widetilde{\rho}_0 \cos \epsilon 
+ A_0(t) \sin \theta_0 \sin \epsilon 
\end{equation}
where $A_{0}(t)$ is given by
\begin{equation}
\label{a0}
A_{0}(t) = - \sin \theta_0 <S_0^z \sin (2h_0t +\phi_0)>
\end{equation}
and $h_0=\sum_j J_{0j}S^z_j$ is the internal field. The strength of the 
perturbation determines the amplitude of the rotation from 
the 0-logical state $(\epsilon=0)$ and 1-logical state 
$(\epsilon=\frac{\pi}{2})$. 
Note that only the 1-logical state component of (\ref{1bitgate}) will evolve
in time and then will be affected by decoherence ( $A_{i}(t)\rightarrow 0$ )
while permanent modification are found in the 0-logical state 
amplitude. A simple oscillating behaviour is 
obtained in the MFA when internal field fluctuations are neglected, i.e. 
$h_i= \sum_j J_{ij}m_j$ and 
$ m_i = \frac{1}{2} \tanh \left(\beta h_i \right) $. It will be shown that a 
finite decoherence time arises as a consequence of non vanishing spin 
correlations $c_{ij} = <S_iS_j>-<S_j><S_j> $ ( neglected in the MFA ).

The 2-qubits gate can be realized by switching on a particles transfer 
between two sites. A simple result is obtained in the case of sites i,j
belonging to two non-interacting adiacent layers of opposite magnetization.
The Quantum coherence of the 2-qubits states associated with sites i and j is
conveniently characterized by the statistical average of the product of 
reduced two sites quantum coherence operators 
$\widetilde{\rho}_{ij}=<(\pi_i-\frac{1}{2})(\pi_j-\frac{1}{2})>$. At 
equilibrium the two sites 0-logical states is characterized as 
\begin{equation}
\widetilde{\rho}_{ij} = <S_i^zS^z_j> \cos \theta_i \cos \theta_j 
\end{equation} 
Note that in the MFA, where spin correlation are neglected, i.e. 
$<S_iS_j>=<S_j><S_j>$ $(i\neq j)$ the two sites reduced QCP is equal to the 
product of one site reduced QCP. The impulsive perturbation hamiltonian 
associated with sites i,j is then 
\begin{equation}
H_I=S_i^+S_j^- +S_j^+S_i^- 
\end{equation}
Where i and j belong to two different non-interacting layers of opposite
magnetization.
In the time evolution of the reduced two sites QCP, following the impulsive 
perturbation we distinguish a constant term and a rotated one associated only 
with $|\uparrow_i,\downarrow_j>$ and $|\downarrow_i,\uparrow_j>$ states.
\begin{equation}
\widetilde{\rho}_{ij}(t) = \widetilde{\rho}_{ij}(0) +
A_{ij} \sin \left( 2 \epsilon \right) \sin \theta_i \sin \theta_j
\end{equation}
where $A_{ij}$ is given by
\begin{equation}
\label{aij}
A_{ij} =  \frac{1}{4} <\left(S_i^z - S^z_j\right) 
\sin\left[\left(h_i-h_j\right)t+\phi_i-\phi_j\right]>
\end{equation}
The impulsive local interaction produces the 1-logical state on two
sites which oscillate with the interaction strenght. The peculiar 
superposition of 0 and 1-logical states is a consequence of the 
quantum entanglement induced by the interaction.
In the MFA limit, the internal fields $h_i$ and $h_j$ become non fluctuating
quantities to be determined self consistently. To evaluate decoherence 
corrections to MFA must be taken into account in the framework of
a perturbation expansion in the interaction range \cite{Georges}.
If the interaction range is small with respect to the system size the 
relaxation toward equilibrium can be seen as a quantum diffusion process wich 
spreads the perturbation over the whole lattice.
We use the technique ref. of\cite{Georges} to calculate an auxiliary free 
energy functional where a single site time dependent field appears
\begin{equation}
\label{functional}
G^{\sigma}(\lambda,t)= \ln \left\{ Tr \left[ exp \left( -\beta H +\sum_k 
(2i \sigma t g_k+\lambda_k) S^z_k\right) \right]\right\}
\end{equation}
Here $\sigma$ is a dicotomic quantity ($ \sigma=\pm $) and  
$g_{k}=J_{0k} $ in the one site ( i=0 ) gate case and 
$g_{k}=(J_{0k}-J_{1k})$ in the two sites ( i=0, j=1 ) gate. In terms of this 
dynamical functional we can express both the one (\ref{a0}) and the two sites 
(\ref{aij}) QCPs 
variations due to one and two sites gates
%\uprule
\begin{equation}
A_0= \frac{i}{2} \sum_{\sigma=\pm}\left. 
\sigma \: exp\left( G^{\sigma}(t)-G^{\sigma}(0)\right)
\frac{\partial G^{\sigma}}{\partial \lambda_0}\right|_{\{ \lambda_j\}=0}
\end{equation}
\begin{equation}
A_{01}= \frac{i}{8} 
\sum_{\sigma=\pm}\left. \sigma\:
 exp\left( G^{\sigma}(t)-G^{\sigma}(0)\right)e^{i\sigma(\phi_0-\phi_1)}
\left( \frac{\partial G^{\sigma}}{\partial \lambda_0} - 
\frac{\partial G^{\sigma}}{\partial \lambda_1} 
\right) \right|_{\{ \lambda_j\}=0}
\end{equation}
%\downrule
As a simple generalization of procedure of ref. \cite{Georges}, we introduce a 
Legendre transform at fixed time and temperature
\begin{equation}
\label{legendre}
F^{\sigma}(m,\beta,t) = G^{\sigma}(\lambda,\beta,t)-\sum_k 
\lambda_k(\beta,t) m_k
\end{equation}
Where $m_k=\frac{\partial G^{\sigma}}{\partial \lambda_0}$. 
The dynamical functional of (\ref{legendre}) is obtained adding an imaginary
time dependent term to the static functional. 
\begin{equation}
\label{dyn}
F^{\sigma}(m,\beta,t)=F(m,\beta)+\sum_k 2 i \sigma t g_k 
\end{equation}
The time independent part of the (\ref{dyn}) $F(m,\beta)$ is given in 
ref. \cite{Georges}. It is conveniently splitted in two terms 
\mbox{ $F(m,\beta)=F_0(m)+F_1(m,\beta)$ }. The first term is associated to 
the non interacting system with fixed magnetization and the latter takes into 
account sites interaction.  
%\uprule
\begin{equation}
F_0(m)= - \sum_{k} \left[ (m_k + \frac{1}{2}) \ln (m_k + \frac{1}{2}) 
+ (\frac{1}{2}-m_k) \ln (\frac{1}{2}-m_k ) \right]
\end{equation}
\begin{equation}
F_1(m,\beta)= + \beta \sum_{lk}J_{lk} m_l m_k 
+ \beta^2 \sum_{lk} J^2_{lk} \left( \frac{1}{4} - \left. m_k \right.^2 
\right)\left( \frac{1}{4} - \left. m_l \right.^2 \right)+ \cdots
\end{equation}
%\downrule
The main point is to study the extremum condition 
$\frac{\partial F^{\sigma}}{\partial m_k}=0$ in the complex plane. 
\begin{equation}
\label{minimo}
\frac{\partial F(m,\beta)}{\partial m_k}= - 2 i \sigma t g_k 
\end{equation}
For short time ( $J_{0k} t \ll 1$ ) we can express the solution of the 
extremum problem as $m_k^{\sigma}(t)=m_k +\delta m_k^\sigma $ where $ m_k $
is the solution at t = 0. Expanding to the first order equation (\ref{minimo}) 
we obtain
\begin{equation}
\label{minimo2}
\sum_l\frac{\partial^2 F(m,\beta)}{\partial m_k \partial m_l}\delta m_l^\sigma 
= - 2 i \sigma t g_k 
\end{equation}
From (\ref{minimo2}) we see that 
$\delta m_k^\sigma = it \sigma \delta\overline{m}_k$ where 
$\delta\overline{m}_k \propto \frac{1}{D}$ and, then, vanishes in the 
MFA limit. The same expansion gives for the dynamical function of 
(\ref{legendre}) 
%\uprule
\begin{equation}
F^{\sigma}(m,\beta,t) \approx F(m,\beta) + \sum_k \left[  2 i \sigma t g_k m_k 
- t^2 \delta\overline{m}_k g_k   
\right]
\end{equation}
%\downrule
From (\ref{minimo2}) we can derive the time dependence of $F(m)$ 
\begin{equation}
F^{\sigma}(m,\beta,t) \approx F(m,\beta) +\sum_k  2 i \sigma t g_k m_k 
-\Gamma t^2 
\end{equation} 
The quantity $2 i t \sigma \sum_k g_k m_k $ is related to the oscillating 
behaviour of the MFA while $\Gamma = \sum_k g_k \delta\overline{m}_k $ is real
 quantity related to the decoherence time.
Because of the stability condition on the extremum of $F(m)$, $\Gamma > 0$. 
If the number of next neighbors of site 0 ( for one bit gate ),
or sites 0 and 1 ( for two bits gate ), which are next neighbors 
each other do not scale with the dimensionality, then 
$\delta\overline{m}_k $ to first order in $\frac{1}{D}$ is equal to
$\delta \overline{m}_k = g_k \left(\frac{1}{4}-m_k^2 \right)$
and $\Gamma$ becomes
\begin{equation}
\Gamma= \sum_k g_k^2 \left(\frac{1}{4}-m_k^2 \right)
\end{equation}
As expected $\Gamma$ is proportional to $\frac{1}{D}$ and vanishes in the 
limit of infinite dimensionality ( MFA limit ). It is important to observe 
that decoherence decreases as magnetization approaches saturation 
$(m^2 \rightarrow \frac{1}{4})$.

In conclusion we studied quantum computation in a spin system at thermal 
equilibrium in the presence of a spin symmetry spontaneous breaking. With 
respect to the conventional approach based on out of equilibrium systems two
main results seem to be of interest.

a) It is possible to define logical states as suitable thermal averages 
instead of quantum states. 

b) Decoherence, due to the conservative dynamics associated to energy operator
wich defines statistical equilibrium, can be treated as a perturbation effect 
with respect to MFA and it is even possible to reduce it at will by increasing
the order parameter. 

We think that the main results of our work should be valid for other systems
which exhibit spontaneous symmetry breaking.

%%%%%%%%%%%%%%%%%%%%%%%%%%%%%%%%%%%%%%%%%%%%%%%%%%%%%%%%%%%%%%%%%%%%%%%%%%%%%%%
%%%%%%%%%%%%%%%%%%%%%%%%%%%%%%%%%%%%%%%%%%%%%%%%%%%%%%%%%%%%%%%%%%%%%%%%%%%%%%%
%%%%%%%%%%%%%%%%%%%%%%%%%%%%%%%%%%%%%%%%%%%%%%%%%%%%%%%%%%%%%%%%%%%%%%%%%%%%%%%
%%%%%%%%%%%%%%%%%%%%%%%%%%%%%%%%%%%%%%%%%%%%%%%%%%%%%%%%%%%%%%%%%%%%%%%%%%%%%%%
%%%%%%%%%%%%%%%%%%%%%%%%%%%%%%%%%%%%%%%%%%%%%%%%%%%%%%%%%%%%%%%%%%%%%%%%%%%%%%%

%
% The bibliography
%

%\end{multicols}


\begin{thebibliography}{99}
\bibitem{nota} It is interesting to note that the order parameter is also 
given by the angular average of $\cos \theta_i$ with respect to $\rho_i$
\end{thebibliography}
\end{document}